\documentclass[conference]{IEEEtran}
\IEEEoverridecommandlockouts
\usepackage{cite}
\usepackage{amsmath,amssymb,amsfonts}
\usepackage{algorithmic}
\usepackage{graphicx}
\usepackage{textcomp}
\usepackage{xcolor}
\usepackage{hyperref}
\usepackage[style=base]{caption}
\usepackage{subcaption}

\usepackage{listings}

\makeatletter
\newcommand{\linebreakand}{%
  \end{@IEEEauthorhalign}
  \hfill\mbox{}\par
  \mbox{}\hfill\begin{@IEEEauthorhalign}
}
\makeatother

\definecolor{code_string}{rgb}{0.533, 0, 0}
\definecolor{code_number}{rgb}{0, 0.533, 0}

\begin{document}
\title{The Service Analysis and Network Diagnosis Data Pipeline

}
%
%

\author{\IEEEauthorblockN{Derek Weitzel}
\IEEEauthorblockA{
\textit{University of Nebraska -- Lincoln}\\
Lincoln, NE \\
dweitzel@unl.edu}
\and
\IEEEauthorblockN{Shawn McKee}
\IEEEauthorblockA{
\textit{University of Michigan}\\
Ann Arbor, MI \\
smckee@umich.edu}
\and
\IEEEauthorblockN{Brian Paul Bockelman}
\IEEEauthorblockA{
\textit{Morgridge Institute of Research}\\
Madison, WI \\
bbockelman@morgridge.org}
\linebreakand
\IEEEauthorblockN{John Thiltges}
\IEEEauthorblockA{
\textit{University of Nebraska -- Lincoln}\\
Lincoln, NE \\
jthiltges@unl.edu}
\and
\IEEEauthorblockN{Marian Babik}
\IEEEauthorblockA{
\textit{European Organisation for Nuclear Research}\\
Geneva, Switzerland \\
Marian.Babik@cern.ch}
\and
\IEEEauthorblockN{Ilija Vukotic}
\IEEEauthorblockA{
\textit{University of Chicago}\\
Chicago, IL \\
ivukotic@uchicago.edu}
}

\maketitle

\begin{abstract}
    Modern network performance monitoring toolkits, such as perfSONAR, take a remarkable number of measurements about the local network environment.  To gain a complete picture of network performance, however, one needs to aggregate data across a large number of endpoints.  The Service Analysis and Network Diagnosis (SAND) data pipeline collects data from diverse sources and ingests these measurements into a message bus.  The message bus allows the project to send the data to multiple consumers, including a tape archive, an Elasticsearch database, and a peer infrastructure at CERN.  In this paper, we explain the architecture and evolution of the SAND data pipeline, the scale of the resulting dataset, and how it supports a wide variety of network analysis applications.
\end{abstract}

\section{Introduction}
\label{intro}

The Research and Education (R\&E) community has a plethora of monitoring available for their networks---from traditional packet counters on routers to netflow-based \cite{rfc3954} data about flows on the local network.  The larger challenge, however, has often been to assemble these individual ``snapshots'' of data into a montage of the entire ecosystem.  The researchers who utilize these networks do not care about the performance of a single network segment but rather the end-to-end performance of their data transfers.  Tracking end-to-end performance is especially acute in the research arena as data flows across ``long fat networks'': that is, the networks often have high data rates (currently, 100's of gigabits across the backbone) and high latencies (often, transcontinental); maintaining acceptable TCP performance is notoriously problematic across such networks.

A key development in assembling the end-to-end picture was the development of the perfSONAR toolkit \cite{ps}.  These toolkits, deployed in relevant locations (such as next to data sources or sinks), could collectively schedule network performance tests and provide data about TCP flows, round-trip time, or packet loss between two sites.  Perhaps more important than the test themselves, perfSONAR helped build interest in observing network performance and an openness in the community to collaborate across site and network boundaries and share data.  Almost any toolkit could initiate tests with another and the data stored in the toolkits were freely accessible.

The Service Analysis and Network Diagnosis (SAND) data pipe\-line builds on the open community by aggregating the test results kept in perfSONAR toolkits and other data sources into central repository of network performance information.  It provides a data bus where test results can be posted and disseminated, a set of ingesters to pull data from remote repositories, and a database where the information is kept online.  The data pipeline, described in this paper, powers a set of visualization tools to help the community build a bird's eye view of the global network.

\section{SAND Communities}

The concept behind the SAND data pipeline is largely driven by the needs of worldwide distributed computing projects; particularly, the Open Science Grid (OSG) \cite{osg} in the US and the Worldwide LHC Computing Grid (WLCG) \cite{wlcg}.

The OSG and the WLCG have been supporting network monitoring activities since 2012, focusing on assisting their users and affiliates on improving their overall network throughput by introducing active monitoring of their networks and providing the ability to test for and identify potential network performance bottlenecks \cite{osg,wlcg}. Two important areas of development were establishing and operating a global network of measurements agents and development and operations of a comprehensive networking monitoring platform which collects and stores the measurements while making them available for further processing.  These agents, based on perfSONAR, are the majority of the inputs to the SAND data pipeline. This structural work has been complemented by several activities that have improved our ability to manage and use both network topology and network metrics to extract clearer understanding of our network problems, locations and bottlenecks via analytics \cite{wlcg-NTW}. 

The WLCG Network Throughput Working Group \cite{wlcg-workinggroup} was established in 2014 to help with operational tasks, such as overseeing the global network of perfSONAR measurement agents based, establishing baseline measurements and performing low-level debugging activities. This has led to the creation of a dedicated network throughput support unit, which has proven capable of successfully coordinating and resolving complex network performance incidents within LHCOPN and LHCONE \cite{lhcone}. However, if is important to note that this group's success requires end-to-end historical network monitoring data.  It is insufficient to only have monitoring of the regional or backbone networks: the Working Group has often found the complete end-to-end network monitoring illuminates otherwise invisible issues between end-site and regional networks or across network provider boundaries.



Networks that connect sites and scientific experiments need to handle ever increasing amounts of data and convey it around the world. Due to the underlying complexity, end-to-end performance depends on a number of components and their operational status anywhere within not only one network, but the sequence of networks the data transits. When transfers are under-performing or errors occur, it is difficult to isolate and correct the source of the problem as local testing by individual network may reveal no problems. While hard failures, such lack of connectivity, are relatively easy to detect and fix, soft failures where a network continues to function but has compromised performance are more subtle.  These soft failures often can only be identified by the active end-to-end measurements against a predefined target and looking for variations in trends across historical data. The OSG and WLCG communities, through the WLCG Network Throughput Working Group, use the SAND data pipeline in order to spot these trends and help address network performance issues.

\section{Related Work}

The SAND data pipeline provides an aggregation mechanism for network test results utilizing a set of end-to-end measurement tools.  However, it's not the first project to attempt to assemble a view of network health; there are a number of tools and projects that attempt to map out the Internet and monitor its performance.  Some are simple tools, such as the ``looking glass'' page on Hurricane Electric's BGP toolkit (\url{https://bgp.he.net/}).  The projects we believe are closest to SAND are PingER, RIPE ATLAS, and NetSage.

Since 1995, the PingER project \cite{pinger}
has been recording measurements of round trip times between the various measurement points and remote sites.  PingER provides a simple, yet broad metric as it only requires remote endpoints to respond to ICMP packets; SAND provides a way to ingest and archive measurements but is agnostic to the underlying test type.

The RIPE ATLAS \cite{ripe} project aims to build a network measurement archive covering the entirety of the Internet.  It provides a small pre-configured network device that takes simple active measurements, including ping, traceroute, DNS, and SSL/TLS probes.  Whereas RIPE targets a broad base of probes, the data ingested by SAND aims more narrowly at R\&E networks (allowing more resource-intensive tests) and application-level performance. 

Finally, the NetSage Measurement Framework \cite{netsage} and corresponding project collects data from a variety of network sources.  Some of the data types---such as perfSONAR endpoints---overlap with SAND, as well as the target community (R\&E networks) and technologies (Elasticsearch).  Comparatively, SAND focuses more on end-site performance and application-level network performance.  Further, we believe the use of a flexible transport bus is critical as it allows records to be moved across multiple data sinks.


\section{Architecture}
\label{sec:architecture}

The SAND data pipeline ultimately is a mechanism to move data from the measurement location through long-term storage and then to visualization.  It is helpful to break down the architecture by the following phases of the pipeline:

\begin{enumerate}
    \item \textbf{Measurement}: Creating the data by measuring the network performance.
    \item \textbf{Data collection}: Periodic collection from multiple measurement sources.  Sends the data to the data bus.
    \item \textbf{Data bus}: Fault tolerant data bus that receives data from the collectors and routes data to multiple ingesters.
    \item \textbf{Data ingestion}: Transform the raw data from the bus to a useful format, enriches the data from other sources, and prepares data for the storage system.
    \item \textbf{Data storage}: Uses Elasticsearch to store data at two sites. 
    \item \textbf{Data access and visualization}: Provides access to the Elasticsearch database and uses data visualization frameworks to display status of the network.
\end{enumerate}

The remainder of the section is organized around these phases, with one subsection for each grouping.  The overall architecture is shown in Figure \ref{fig-4}.

\subsection{Measurement}
\label{sec:measurement}
While SAND can handle a variety of data types, there is a strong focus on integration with the perfSONAR toolkit due to its popularity across end sites in the OSG and WLCG communities.
perfSONAR provides a means to schedule active network measurements for metrics such as latency and throughput between two instances.  Each data type is measured by a dedicated tool within the toolkit; each tool's output, along with the test conditions, are recorded in an ``measurement archive'' located at the toolkit originating the test.

The standard set of network metrics supported by perfSONAR are: 

\begin{itemize}
    \item \textbf{Latency}: Latency of communications between two hosts can be measured through either \emph{one-way and two-way active measurement protocols} (OWAMP/TWAMP)\cite{rfc4656}. Note that latency measurements send 10 Hz of timestamped packets (by default) and therefore additionally provide a measurement of packet loss, since each minute, perfSONAR counts how many of the expected 600 packets actually arrive at the destination.  Having a well synchronized clock is critical for this measurement.
    \item \textbf{Throughput}: Typically, perfSONAR measures throughput via the performance of a single TCP flow.  Three different tools are offered: \emph{iperf3, iperf2 and nuttcp}.  The most common is \emph{iperf3}, which can perform memory to memory tests over UDP or TCP and reports TCP retransmits and size of congestion window.  The non-default test configurations (e.g., UDP) are often used by operators for troubleshooting specific issues.
    \item \textbf{Network Path}: Measured by the \emph{traceroute} or \emph{tracepath} tools, the network path is an attempt to discover all layer 3 routers between two toolkits.  Currently, \emph{traceoute} is used even though \emph{tracepath} would be preferred for the additional information it can provide.  The problem is that while \emph{tracepath} can additionally determine \emph{maximum transmission unit} (MTU) along the path, we have numerous cases where the tool fails so successfully measure a path.
\end{itemize}

The measurement archive in each perfSONAR instance used to store the resulting measurements is implemented by the esmond \cite{perfsonar-esmond} service.  In addition to an internal database, the esmond service exposes a HTTP API to enable remote queries.

One of the critical components of perfSONAR that we rely upon is {\bf PWA} (pSConfig Web Administrator)\cite{PWA}.  This tool allows us to centrally define and control which hosts and measurements are run amongst our global deployment of perfSONAR toolkits. PWA gathers information about which toolkits are available from a combination of the toolkit lookup service\cite{pS-lookup} (a global service registry for perfSONAR) and our OSG/WLCG perfSONAR registration information\cite{OSG-topology,GOCDB} and provides an administrator friendly, secured web interface to organize those instances into testing meshes of toolkits.   In addition, it supports configuration (via pSConfig\cite{psconfig-website}) of the various testing tools and their associated test parameters. As we will discuss in Section \ref{sec:data_bus}, we have leveraged this tool to enable secure writing directly to our message bus. A powerful feature of PWA is that it becomes the configuration reference for all associated toolkits via a one-time configuration on each toolkit as they are deployed. The toolkits have a unique URL from the PWA server that provides the complete JSON configuration for the toolkit.  As the mesh or tests are changed, the toolkit's configuration is updated the next time they access its unique URL, typically once per hour.

Beyond perfSONAR, the OSG collects network measurements by instrumenting TCP flows for production data transfers in support of its distributed high throughput computing (dHTC) workflows.
OSG users submit batch jobs to a central OSG access point; the jobs often require significant data movement from the submit host to the worker nodes at remote sites in order to complete, meaning end-to-end network performance is relevant.
The TCP flows monitored by submit hosts measure the network conditions between the worker nodes and submit hosts during file transfers (generally, these occur during the job startup and completion, not continuously during the job lifetime).

Given OSG is present at about 100 sites within the US (and jobs can go to international locations as well), these data transfers can capture aspects of the end-to-end path that might be untested by perfSONAR.  For example, in the OSG, worker nodes can be behind a firewall or a NAT device and, in such cases, perfSONAR would often be connected at the network edge and would not be measuring the same network path.  While a network engineer might normally ignore these ``middleboxes,'' their existence can dramatically affect observed job data transfer throughput.

OSG leverages HTCondor \cite{htcondor} to execute its workflows and HTCondor is configured to output TCP statistics for each job instance's data transfer connections between the submit host and the worker node.  These TCP statistics are provided by the Linux kernel and include the number of loss packets, bytes transferred and number of TCP reordering events.  These statistics are written to a log file by the HTCondor submit host; the file is subsequently parsed by an instance of Filebeat\cite{filebeat}, a log aggregation platform. 



We also gather SNMP data from ESnet, gathering SNMP statistics on interface traffic, errors and discards from over 14,000 interfaces in ESnet. While potentially a very powerful data set for understanding near real-time and historical use of the ESnet network, we still face a significant challenge in correlating a traceroute involving ESnet routers with the specific interfaces we have data from.   The problem is that ESnet using internal traffic engineering that hides the actual physical path in their network, preventing us from easily determining which interface is associated with a given traceroute involving ESnet.

Another common data flow on OSG is between running jobs and data storage services.  For this, SAND relies on measurements collected by the XRootD \cite{dorigo2005xrootd} server; XRootD utilizes the same Linux kernel interface to instrument the TCP connections.  For the SAND pipeline, the XRootD developers created a plugin interface and we created a plugin \cite{xrootd_tcp_stats} to export TCP metrics for each connection.  The plugin receives the same TCP statistics data as HTCondor and converts the data to a JSON data structure which is then sent over UDP to a configured central monitoring collector (see Section \ref{sec:data_collection_xfer}).


\subsection{Data Collection}
\label{sec:datacollection}
\subsubsection{Polling perfSONAR Toolkits}
While perfSONAR measurements are stored in measurement archives at several hundred toolkit instances around the world, SAND needs a ``collection'' layer to move the data to a single archive.  Toward this goal, a central collector \cite{ps_collector} was developed to discover relevant instances, query each for the historical data of interest, and sends the data to a message bus.  The central collector is a \textit{poll-based} application: it periodically queries each of the perfSONAR instances of interest for the latest data.

The collector is a Python application that uses the esmond \cite{perfsonar-esmond} HTTP API to retrieve measurements from each toolkit.  Since the toolkits are distributed worldwide and have varying levels of performance, the time required to retrieve data may be significant.  To speed up the overall retrieval rate, the collector queries many toolkits in parallel; 200 processes are started to simultaneously query up to 200 toolkits.  When the collector first starts, it uses all 200 processes to query every toolkit in the configured meshes and then schedules another query for the toolkit every 5 minutes (with some random delays on the order of a minute to avoid synchronizing the load put on the remote toolkits).

Each process is responsible for querying a single toolkit at a time; as SAND only needs to ingest the data from the last run, a simple state file is kept to determine the last recorded data and the esmond queries are adjusted accordingly.  As results are returned, the collector pushes the data onto a message bus and updates the state file.

\begin{figure*}[htb]
\centering
\begin{subfigure}[b]{0.47\textwidth}
  \centering
  \includegraphics[width=\textwidth]{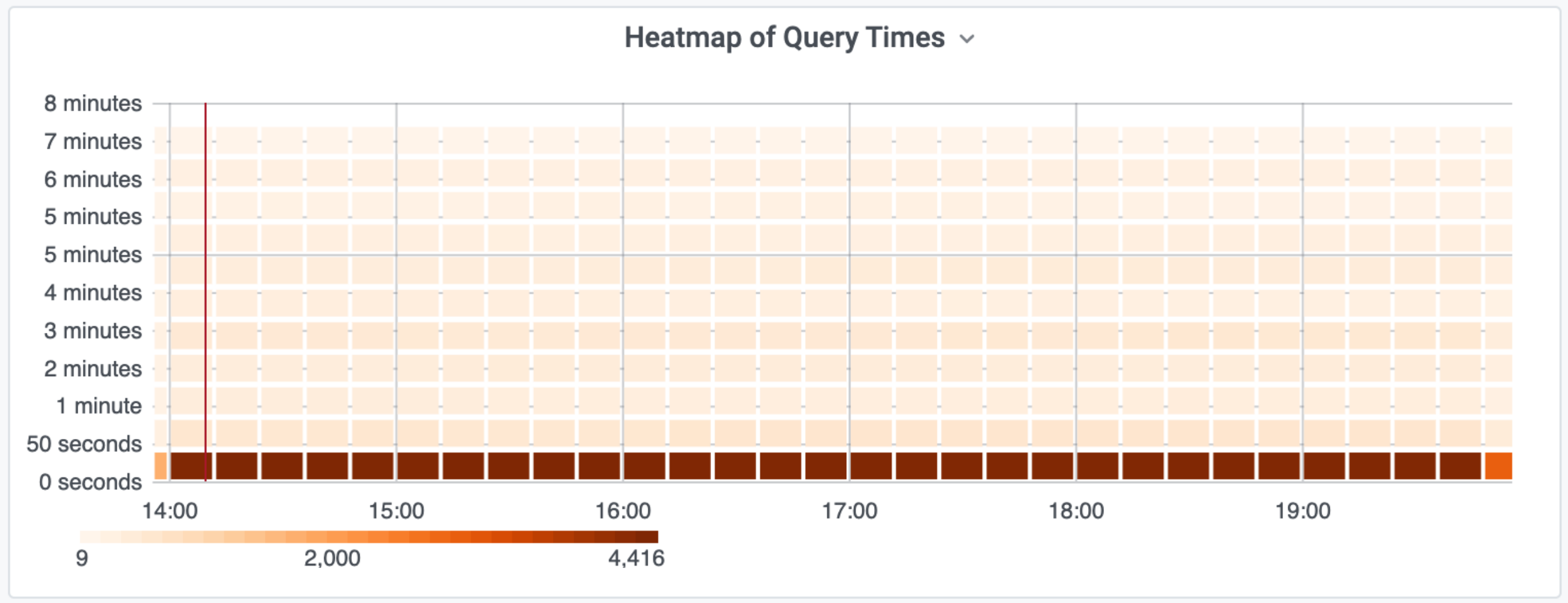}
  \caption{Heat map of query times per toolkit} 
  \label{fig:collectorheatmap}       
\end{subfigure}
\hfill
\begin{subfigure}[b]{0.47\textwidth}
  \centering
  \includegraphics[width=\textwidth]{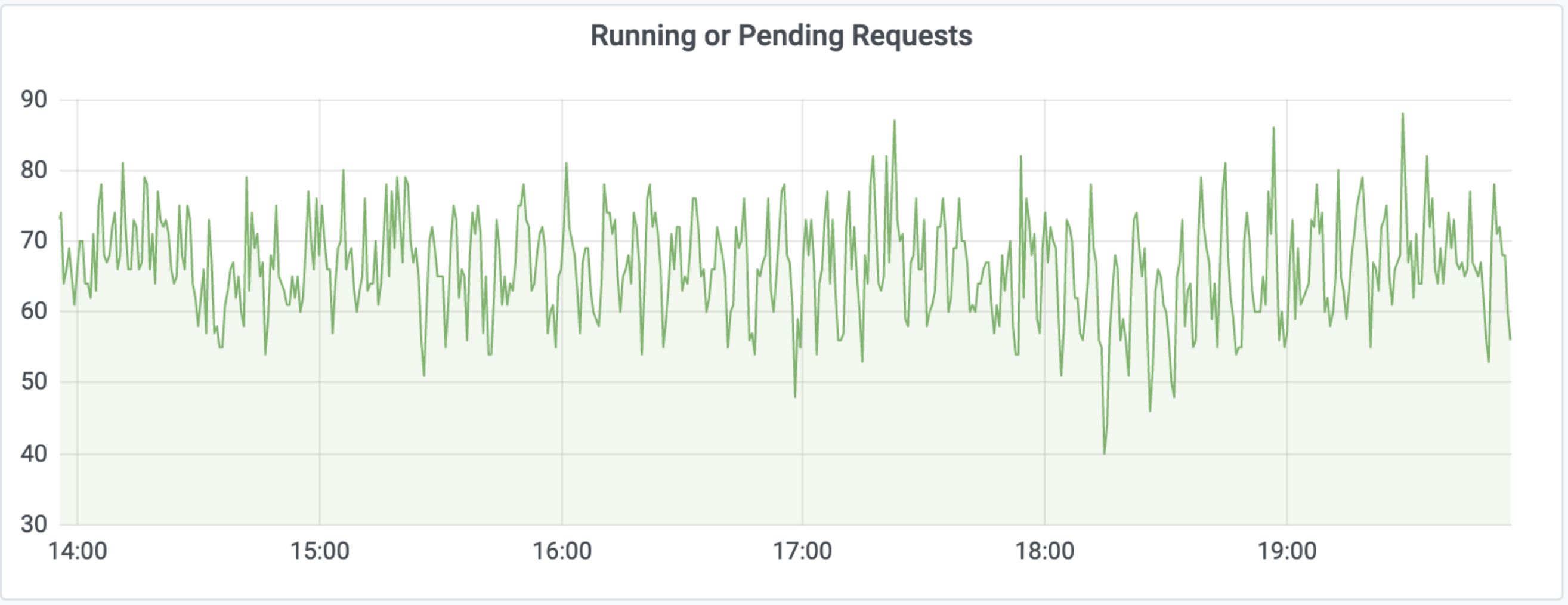}
  \caption{Number of running or pending queries to toolkits} 
  \label{fig:runningpending}       
\end{subfigure}
\caption{Collector Monitoring Grafana dashboard}
\label{fig:collectormonitoring}
\end{figure*}

The collector uses Prometheus \cite{turnbull2018monitoring} to provide monitoring of the queries.  A Grafana \cite{grafana-website} dashboard was created to visualize the status.  The graphs shown in Figure \ref{fig:collectormonitoring} are used to monitor the collector.  Figure \ref{fig:collectorheatmap} is a heat map of queries to toolkits. The vast majority of the queries to toolkits take less than 1 minute to retrieve data.  The line graph in Figure \ref{fig:runningpending} shows the number of toolkits simultanously being queried.  In the current system, at steady state, between 60-80 toolkits are queried at a time by the central collector.

\subsubsection{Direct Push from perfSONAR Toolkits}
\label{sec:directpush}
While polling the remote archives has functioned for several years, there is inherently high latency (at busy times, 15 minutes) before the data is recorded centrally and it requires each toolkit to expose its HTTP API to the Internet.
To tackle these issues, we are transitioning the data collection infrastructure so the toolkit instances can push data directly to the data bus.

Switching from polling to pushing means the toolkit must be authorized to write to the message bus described in Section \ref{sec:data_bus}.
For SAND, the data collection authentication and authorization infrastructure is built on JSON Web Tokens (JWT) \cite{jwt-rfc}.  The JWT is a bearer token and provides a verifiable credential that establishes the bearer's identity and embeds the the authorization permissions for the bearer (the toolkit).
While a powerful mechanism---and supported by the RabbitMQ implementation of the data bus---an automated mechanism is needed to distribute the token to the toolkit instance.  The alternative of requesting each administrator to hand configure a credential would not scale well for our globally distributed set of 100's of toolkits and would additionally dictate assigning much longer-lived tokens.

Luckily, all our toolkits utilize a central configuration service, PWA, previously described.  When the toolkit instance requests its configuration from the OSG's configuration generator, a plugin checks the requesting host IP and the instance configuration requested.  If the requested configuration matches the requested IP, a JWT is created with appropriate permissions to send messages to the RabbitMQ message bus and signed with a private key.  The message bus utilizes RabbitMQ's OAuth2 authorization backend \cite{rabbitmq-oauth} and includes the signer's public key which allows RabbitMQ to verify the JWT and allow the permissions described in the JWT.  Note this token distribution mechanism is ultimately equivalent to host-based security: an attacker capable of spoofing source IP addresses could get a token to upload falsified data.  Given the data value and impact, we view this as an acceptable risk; however, a next phase of the project may leverage the toolkit's host certificate to provide stronger authentication for the token request.


The data pushed from perfSONAR are raw records from the measurement tools whereas the polled data is from esmond (which performs record normalization).  For example, in the pushed data, sources and destination are represented by their IPs instead of by hostnames as they are in the esmond data.  Timing measurements, such as the RTT, for a latency test are represented in the pushed data as ISO 8601 format (\texttt{PT0.0005S}) compared to a float (\texttt{0.0005}) in the pulled data.  This requires data normalization in later steps of the pipeline in order to detect duplicate data.

Currently, depending on the perfSONAR version, not all toolkits can push data, requiring SAND to run the central collector and resulting in a mixture of pushed and polled data.  Accordingly, the collector periodically checks to see if a given toolkit has begun to push data and stops poll-based data collection.  This reduces the volume of duplicate data and to reduces the load on the toolkit from SAND.

\begin{figure}[th]
\centering
\includegraphics[width=0.47\textwidth]{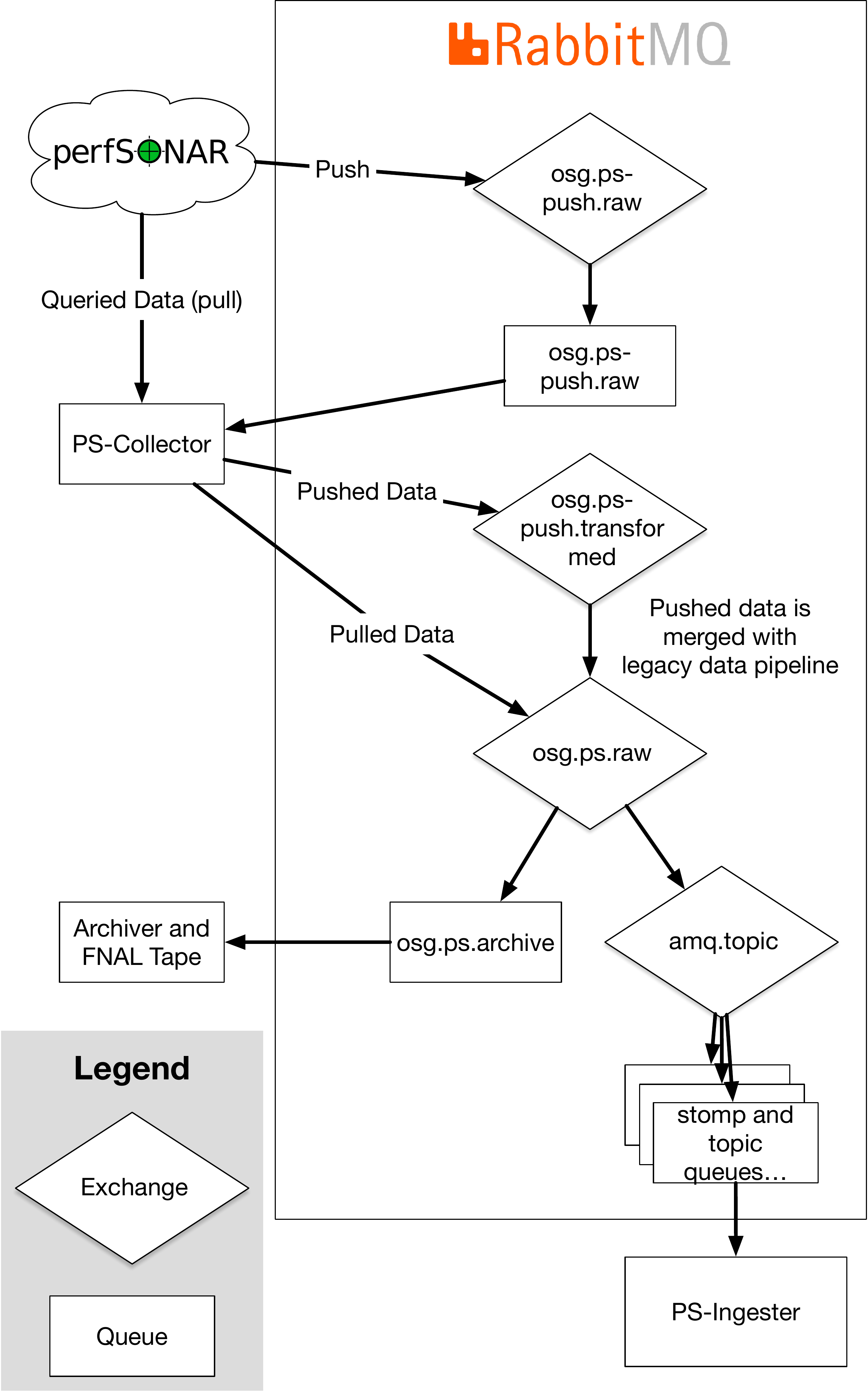}
\caption{Message Bus configuration for the perfSONAR data pipeline} 
\label{fig:perfsonarmessagebus}       
\end{figure}

Figure \ref{fig:perfsonarmessagebus} shows a diagram of the data flow through the message bus.  In the legacy pipeline, the \texttt{ps-collector} process pulls data from perfSONAR toolkits using the esmond API.  In the new push data pipeline, tookits push data directly to the RabbitMQ message bus.  Once the data is pushed to the message bus, it is routed to the collector for normalization.  As both the polled and pushed data is consumed by the collector component, the pipeline is able to detect when a toolkit begins pushing data to the message bus and polling can stop.

\subsubsection{Job and Data Transfer}
\label{sec:data_collection_xfer}
The HTCondor TCP metrics are collected in a process similar to the perfSONAR data.  The data is originally written by HTCondor into a logfile on the submit host and parsed by a Filebeat daemon on the submit host.  The data is sent to a \textit{receiver} \cite{htcondor_xfer_parser}, responsible for performing authentication checks with the client's SSL certificate, providing backoff feedback to the Filebeat clients through the lumberjack protocol, and forwarding the messages to the message bus for eventual ingestion.  Unlike perfSONAR, which required a custom collector, the HTCondor receiver is specially-configured Logstash daemon \cite{turnbull2014logstash} running inside a container.

XRootD sends its TCP networking monitoring data over a UDP stream, serialized in a custom binary format, to a central collector.  The OSG parses this file read and write data with the XRootD Monitor Collector \cite{xrootd_monitoring_collector}; for SAND, this central collector was modified to additionally parse the JSON records containing the TCP metrics.

\subsection{Data Bus}
\label{sec:data_bus}

Whether the data is collected from perfSONAR, HTCondor, or XRootD, it ultimately is sent to a common message bus implemented by RabbitMQ.  The RabbitMQ message bus reliably transmits and routes the messages from multiple producers to multiple consumers (allowing multiple entities to subscribe to the data flows).  Messages can be sent to an \textit{exchange} where the data can be routed internally to different exchanges or \textit{queues} where consumers can read it.  The message bus used for the SAND data pipeline is shared with other OSG use cases but the SAND data pipeline consumes the majority of the bus resources.  The perfSONAR data flow uses 26 queues and, across consumers and producers, has about 225 simultaneous connections in support of daily  activities.  The majority of these are from the \texttt{ps-collector} which, as noted in Section \ref{sec:datacollection}, uses 200 processes to read pull data from toolkits.  Each of those 200 processes uses a separate, long lived, connection to the message bus.

Using a message bus such as RabbitMQ gives the SAND pipeline flexibility to move and reconfigure producers and consumers.  In the lifetime of the project, the perfSONAR data collectors have been migrated between physical sites deploying them as RPMs, as Docker containers, and most recently, deployed in a Kubernetes cluster.  The message bus architecture provides flexibility in the network location of these components in a way a client / server architecture could not.

Eighteen of the queues for the perfSONAR data pipeline are for the collector, nine for each of the two analytic platforms described in \ref{sec:datastorage}.  The queues guarantee delivery of the messages from the producer to the consumer as long as there is disk or memory space to buffer the data.  If a consumer experiences a downtime, the queues will save the undelivered messages.  The internal monitoring systems watch over the length of the queues and alert the operations team if they've gone over preset thresholds.

Figure \ref{fig:perfsonarmessagebus} shows the message bus configuration for the perfSONAR data pipeline.  Section \ref{sec:datacollection} describes how the perfSONAR data is pushed directly to the message bus, routed back, and transformed.  Currently, the pushed data is also routed through a separate exchange from the pulled data allowing the operations team to separately store and debug issues with the pushed data.  Both streams are merged into the \texttt{osg.ps.raw} exchange where the data is duplicated and sent to the \texttt{osg.ps.archive} queue to be stored in tape, or the \texttt{amq.topic} exchange.

The \texttt{amq.topic} exchange is a special exchange in the RabbitMQ message bus.  It is a \texttt{topic} exchange in which each message is tagged with a topic key that determines how the message is routed to the queues.  For example, the \texttt{stomp} \cite{stomp-website} queues subscribe to a specific topic and will only receive messages tagged with that topic key STOMP is a popular messaging protocol implemented in several programming languages and other data buses.  Topic keys are organized by the data type; for example, messages with the \texttt{perfsonar.raw.throughput} topic key contain throughput measurements and the \texttt{perfsonar.raw.packet-trace} key indicates traceroute measurements.

\lstdefinestyle{mystyle}{
    basicstyle=\ttfamily\footnotesize,
    breakatwhitespace=false,         
    breaklines=true,                 
    captionpos=b,                    
    keepspaces=true,                 
    numbers=left,                    
    numbersep=5pt,                  
    showspaces=false,                
    showstringspaces=false,
    showtabs=false,                  
    tabsize=2
}

\lstset{style=mystyle}

\begin{lstlisting}[caption=Decoded payload of a JWT token used for for perfSONAR toolkit authentication,float=ht,label=lst:rabbitjwt,escapeinside={<@}{@>},basicstyle={\ttfamily}]
{
  "scope"    : <@\textcolor{code_string}{"rabbit\_server.write:}@> <@\textcolor{code_string}{osg-nma/osg.ps-push.raw/perfsonar.raw.*"}@>,
  "exp"      : <@\textcolor{code_number}{1618444800}@>,
  "aud"      : <@\textcolor{code_string}{"rabbit\_server"}@>,
  "sub"      : <@\textcolor{code_string}{"ps.example.edu"}@>,
  "client_id": <@\textcolor{code_string}{"ps.example.edu"}@>
}
\end{lstlisting}

To secure the bus sufficiently to allow direct pushes from remote perfSONAR toolkits, the message bus is configured to use JWTs for client authentication and authorization.  The JWTs assigned, created by the plugin described in Section \ref{sec:directpush}, have a finite expiration time and contain access control lists of authorized actions (in JWT terminology, these are \textit{scopes}).  Currently, toolkits are only allow write to the \texttt{osg.ps-push.raw} exchange.  The JWT has the format shown in Listing \ref{lst:rabbitjwt}.  The \texttt{scope} claim defines the authorizations that the holder of the token is allowed to performed.  The Listing \ref{lst:rabbitjwt} example allows the the holder to \texttt{write} messages to a server named \texttt{rabbit\_server}, to the virtual host \texttt{osg-nma}, to the exchange \texttt{osg.ps-push.raw}, only messages with the topic key matching the regular expression \texttt{perfsonar.raw.\*}.  The \texttt{aud} (audience) specifies that this token should only be accepted at a server named \texttt{rabbit\_server}.  The \texttt{sub} (subject) and \texttt{client\_id} specifies that this token was issued to the perfSONAR toolkit at \texttt{ps.example.edu}.  The subject is used solely as a traceability mechanism, as RabbitMQ server does not utilize it for authorization decisions.

The HTCondor job transfer statistics are first gathered by a centrally-operated Logstash receiver.  The receiver verifies the client authorization and forwards the data unmodified to the message bus.  HTCondor collectors subscribe to the message bus, receiving the raw data so they can be enriched before being stored into a database.

Note the conscious design decision for the message bus to separate raw data into different queues from processed and enriched data.  This will allow the raw data to go to a separate archive system.  Not only does the raw data archive provide disaster recovery capabilities but allows the data to be ``replayed'' if the processing pipeline or data enrichment changes in the future.

\begin{figure}[th]
\centering
\includegraphics[width=0.47\textwidth]{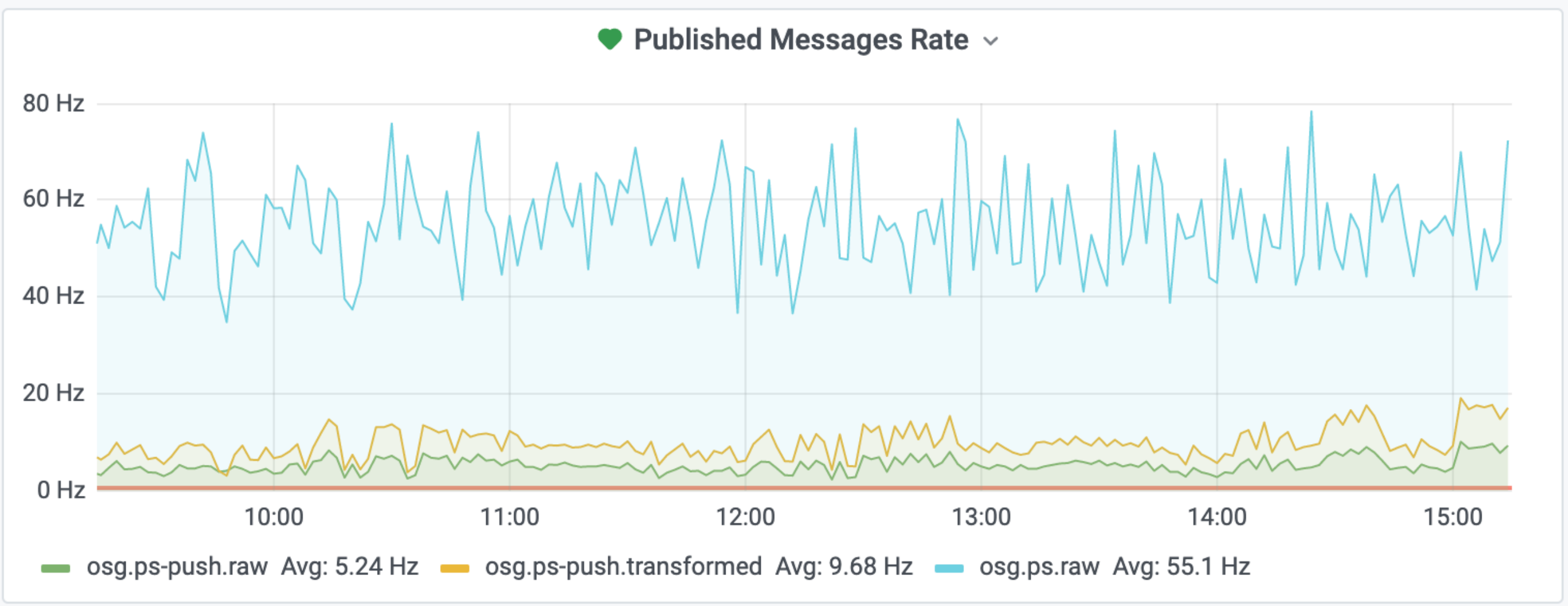}
\caption{Message Bus Grafana monitoring} 
\label{fig:messagebusgrafanamonitoring}       
\end{figure}

\subsubsection{Monitoring} SAND utilizes Prometheus and Grafana to monitor the message bus.  Figure \ref{fig:messagebusgrafanamonitoring} shows the message publishing rate to the three exchanges described in Figure \ref{fig:perfsonarmessagebus}, \texttt{osg.ps-push.raw}, \texttt{osg.ps-push.transformed}, and \texttt{osg.ps.raw}.  The \texttt{osg.ps.raw} is the combination of both the pushed and the pulled data.  As the transition to pushing data directly to the message bus is ongoing, the monitoring shows the pushed data is significantly less than the pulled data.  The monitoring alerts us if the published message rates are either higher than normal or have dropped to zero.

\subsection{Data Ingestion}
At the data ingester phase, dedicated daemons consume raw records from the message bus, enrich the data with attributes from other sources, and prepare and store the data into storage.  The nine perfSONAR ingesters \cite{ps_ingest} add site information and calculate some summary statistics for each record.  Four of the ingesters process status and metadata information, while the other five process latency, packet loss, retransmits, throughput, and traceroute information.

When the perfSONAR ingester starts, it downloads site and topology information from public databases at CERN and the OSG.  Each ingester adds the site information to the records and flatten the incoming data structure to remove nested attributes.  The throughput and packet loss ingesters do not modify the measurements and simply pass the data through.  In contrast, the traceroute ingester reformats the list of IPs along the route and annotates each route segment with the corresponding autonomous system (AS) numbers and RTT.  It further annotates the record if the traceroute record successfully mapped the route to the desired destination.

For each record, the ingesters generate a record ID used by the storage system to detect duplicate records (duplicate records are expected to occur at a low rate, for example, when the poll-based and push-based data collection overlaps for a single toolkit).  The ID is generated by concatenating the test timestamp, source, destination, and test type.  The same ID is generated for both pushed and pulled data; we have verified that, with this unique ID, the pushed data can be overwritten by pulled data from the same toolkit instance.



The ingester for the HTCondor TCP metrics enrich and tag the incoming raw data.  The raw data is a single log line from the HTCondor submission host.  It includes the date, time, and key value pairs of TCP statistics.  The ingester is a Logstash Docker container that with a custom configuration that performs the following actions:

\begin{enumerate}
    \item Tag the record as either a file upload or download
    \item Parse the timestamp and format to it for ingestion into Elasticsearch
    \item Convert values in the key value pairs to their native representation, either floats, integers or leave as strings
    \item Mark the latitude and longitude of the destination, which should be the worker node, using GeoIP
    \item Resolve the IP address of the destination, and add a field with only the domain.  This is useful if we want to group by an institution
    \item Generate the unique ID
    \item Write the enriched record to the Elasticsearch database
\end{enumerate}


Similar to the perfSONAR data, a unique ID is required for each record.  For the HTCondor data, the ID is generated by concatenating the full log line as well as the hostname of the reporting HTCondor host.

Both the perfSONAR and HTCondor ingesters insert the enriched records directly into Elasticsearch using its HTTP API.

\subsection{Data Storage}
\label{sec:datastorage}
The SAND data pipeline aims to provide for easy querying of individual data records (as opposed to summarized information), low maintenance, and rapid ingestion.  The project choose Elasticsearch \cite{gormley2015elasticsearch}, a document-oriented-database, for the storage of the network metrics.  The data ingesters  write their transformed data directly to an Elasticsearch database.  Currently, there are two separate ElasticSearch databases that read from the message bus; one, located at University of Nebraska--Lincoln is optimized for large capacity and allowing queries (potentially slower) across the entire historical dataset.  The second Elasticsearch cluster, at University of Chicago, is powered entirely with SSDs and provides quicker query responses (but holds less data).


In addition to the two Elasticsearch databases, SAND sends a copy of all raw perfSONAR data to a tape archive at the Fermi National Accelerator Laboratory.  A RabbitMQ client listens on the \texttt{osg.ps.archive} exchange of the message bus, as shown in Figure \ref{fig:perfsonarmessagebus}.  The archiver \cite{gracc_archive} appends each message as a file in a ``tar'' archive (which is then compressed).  Every day at midnight, the current archive file is closed and a new one is created.  A separate process periodically uploads the closed archive file to the archive system.

The raw data archive has proven to be a useful resource.  In 2019, the operations team discovered an issue with the original ID generation used for duplicate detection.  Additionally, after performing some basic analysis of the traceroute data, the team realized an additional three attributes were needed from enrichment during ingestion.  Instead of updating the records in the database, the team was able to perform a replay of a full year of raw data from the archive through the data ingestion step.

\begin{figure}[th]
\centering
\includegraphics[width=0.47\textwidth]{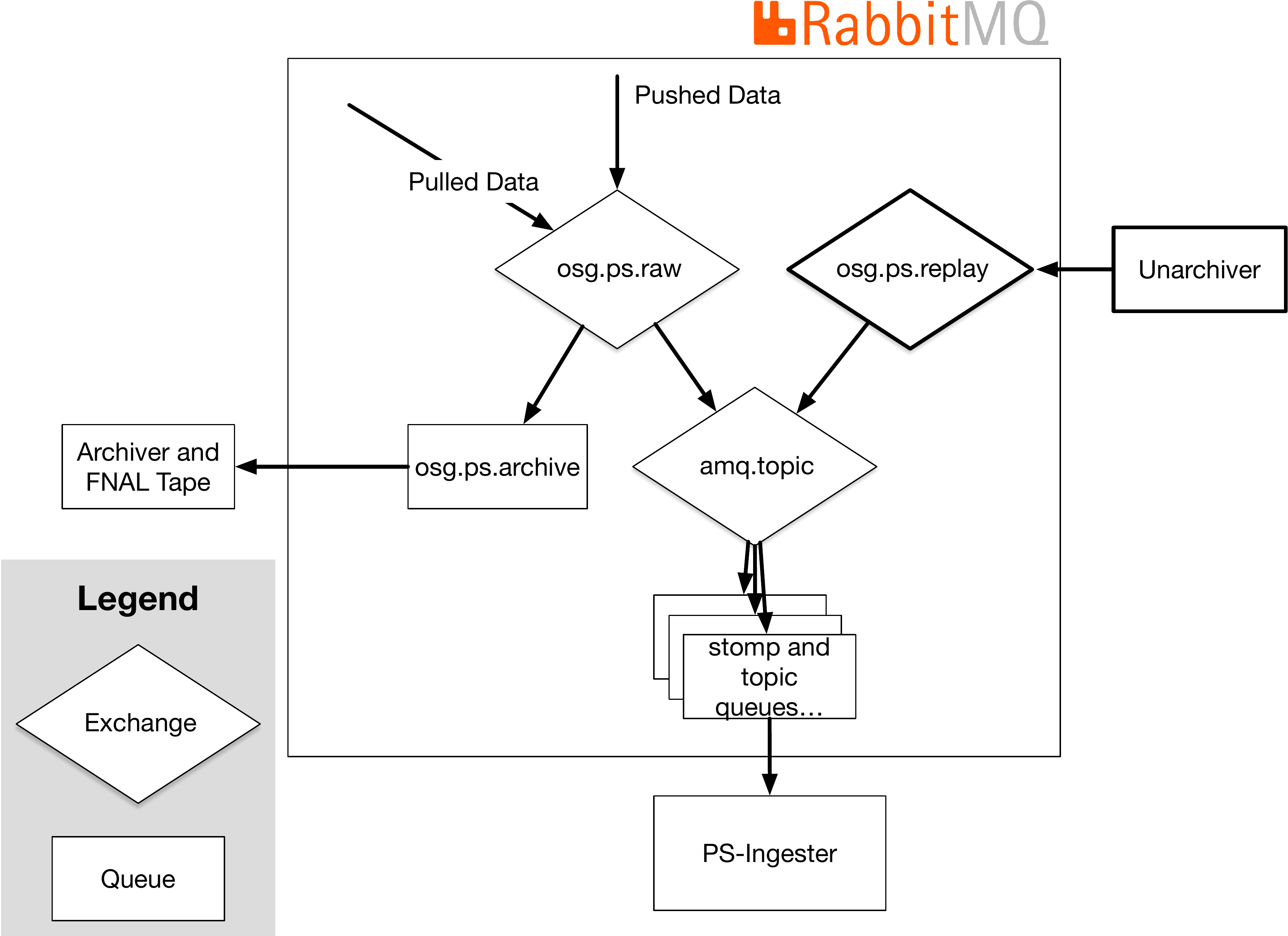}
\caption{Message Bus configuration for tape replay} 
\label{fig:messagebustapereplay}       
\end{figure}

Figure \ref{fig:messagebustapereplay} shows the message bus configuration during the tape replay.  New measurements continued to flow through the production pipeline to the database and tape.  The replayed data was written to an \texttt{osg.ps.replay} exchange and sent to the \texttt{amq.topic} to run through the ingesters again; note the replayed data bypassed the \texttt{osg.ps.raw} exchange in order to avoid being written to tape again.  The message bus gave us the flexibility to route the messages while continuing to take production data.  Overall, the system was able to replay roughly 300 million records in 4.5 days.

\subsection{Data Access and Visualization}
\label{sec:data_access_viz}
One of the most powerful features of our pipeline is the ability to easily access and query the complex dataset.   By using Elasticsearch as the primary storage, users are able to leverage a vibrant ecosystem of tools that integrate with the database; this enables users to quickly select and visualize information of interest.  

\begin{figure}[ht]
\centering
\fbox{\includegraphics[width=0.45\textwidth]{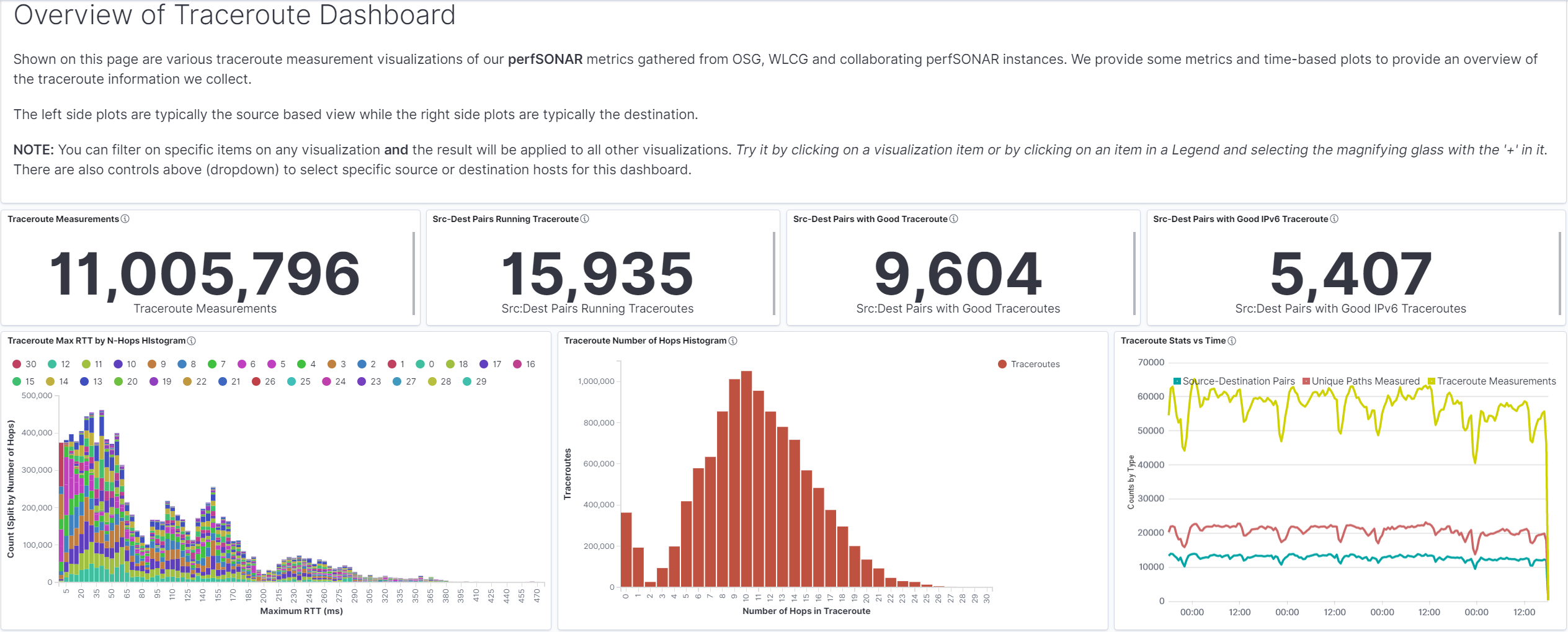}}
\caption{Kibana traceroute dashboard, showing the first traceroute metric visualizations for a 4 day period.} 
\label{fig:kibana-tr}       
\end{figure}

A Kibana web interface\cite{Kibana-MWT2} is available for guest users, though such users are unable to save any visualizations, queries or dashboards that they create. It is also possible for users needing specific visualizations to get an account on the platform, which would allow them to create persistent visualizations, or work with the SAND team to create such visualizations. 

The SAND team has developed simple Kibana dashboards that provide details on packet loss, one-way delay, network throughput and network topology, as well as a dashboard that tracks important characteristics of our toolkits and host systems.  The project's focus is to make the dataset easily accessible; we try to engage external researchers to build custom visualizations from the data.  The Kibana interface allows users to easily query, filter and visualize any of the Elasticsearch data of interest.

\begin{figure}[ht]
\centering
\fbox{\includegraphics[width=0.45\textwidth]{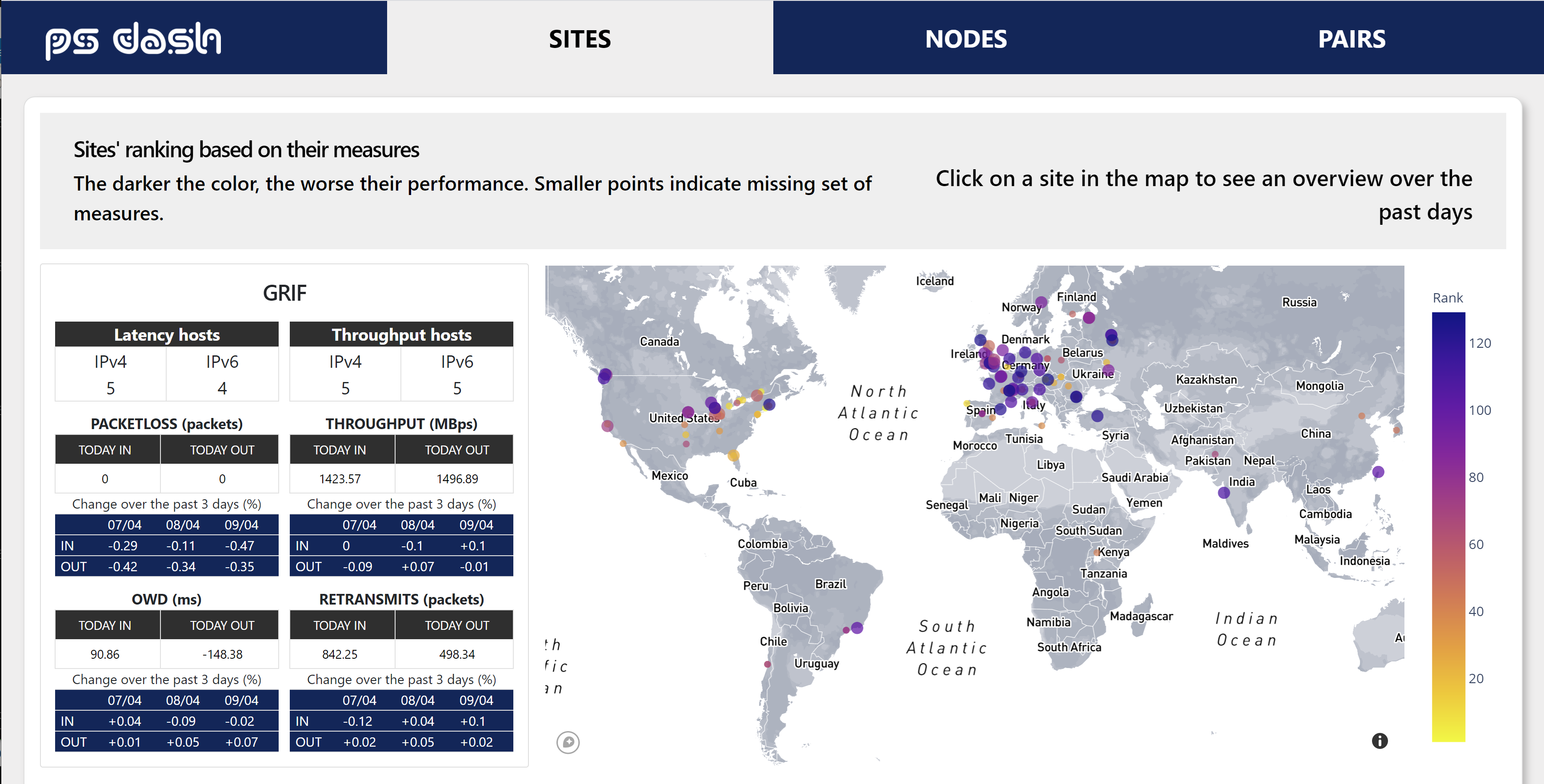}}
\caption{The pS-Dash web interface showing toolkit sites on a global map with darker colors indicating more problematic measurement results.  Selecting any instance provides summary graphs for that site.} 
\label{fig:ps-dash-front}       
\end{figure}

In addition, one of our students has developed a Plotly\cite{plotly}  dashboard called pS-Dash\cite{ps-dash} running on our Analytics Platform that analyzes and visualizes some of our network metrics.  The starting interface is shown in Fig \ref{fig:ps-dash-front} and provides a quick overview of sites color coded based upon how problematic the associated network measurements are for each site.   

\begin{figure}[ht]
\centering
\fbox{\includegraphics[width=0.45\textwidth]{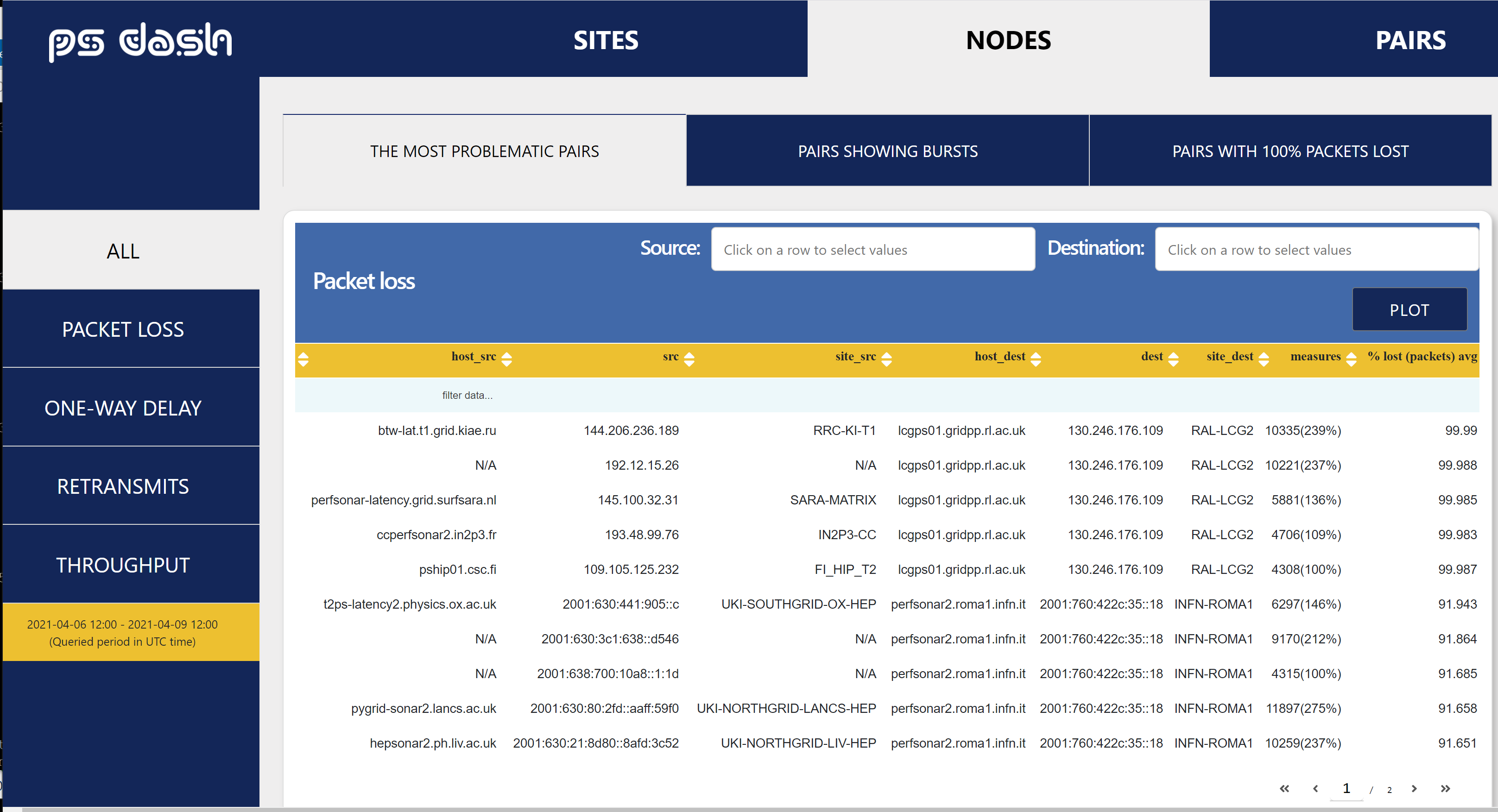}}
\caption{The nodes tab on the pS-Dash web interface showing toolkit pairs with the most problematic measurements.} 
\label{fig:ps-dash-problems}       
\end{figure}

\begin{figure}[ht]
\centering
\fbox{\includegraphics[width=0.45\textwidth]{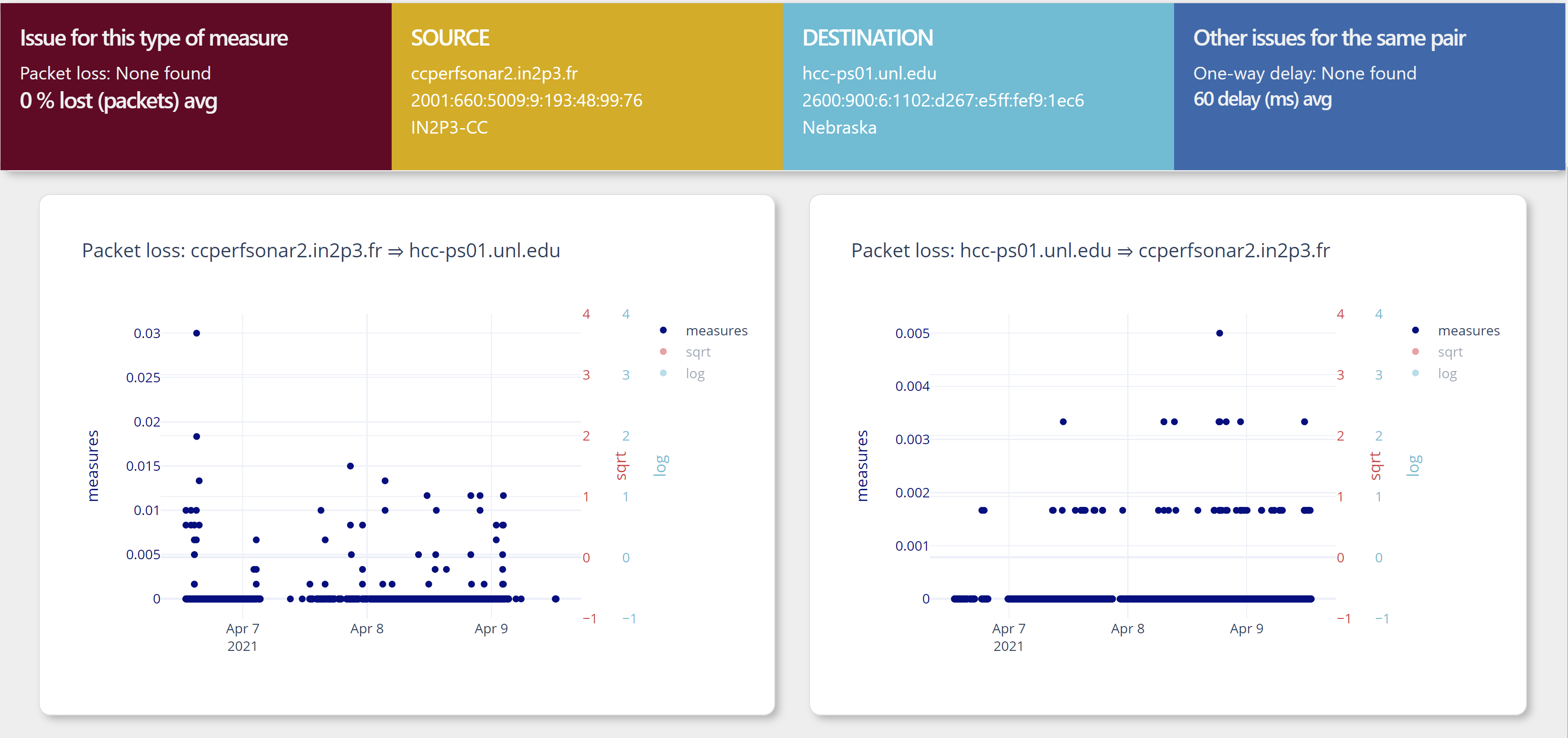}}
\caption{Details on selected pairs in ps-dash can be plotted.} 
\label{fig:ps-dash-select-pairs}       
\end{figure}

The ps-dash interface has tabs that provide ranked views of various problems like high packet loss, large one-way delay variations, large amount of retransmits or low throughput, as shown in Fig \ref{fig:ps-dash-problems}.  Specific toolkit pairs can be selected and the results plotted, as shown in Fig \ref{fig:ps-dash-select-pairs}.  This tool is still under development and is our primary test-bed for creating processing elements to identify specific problems which can then be alerted on.

\begin{figure}[ht]
\centering
\fbox{\includegraphics[width=0.45\textwidth]{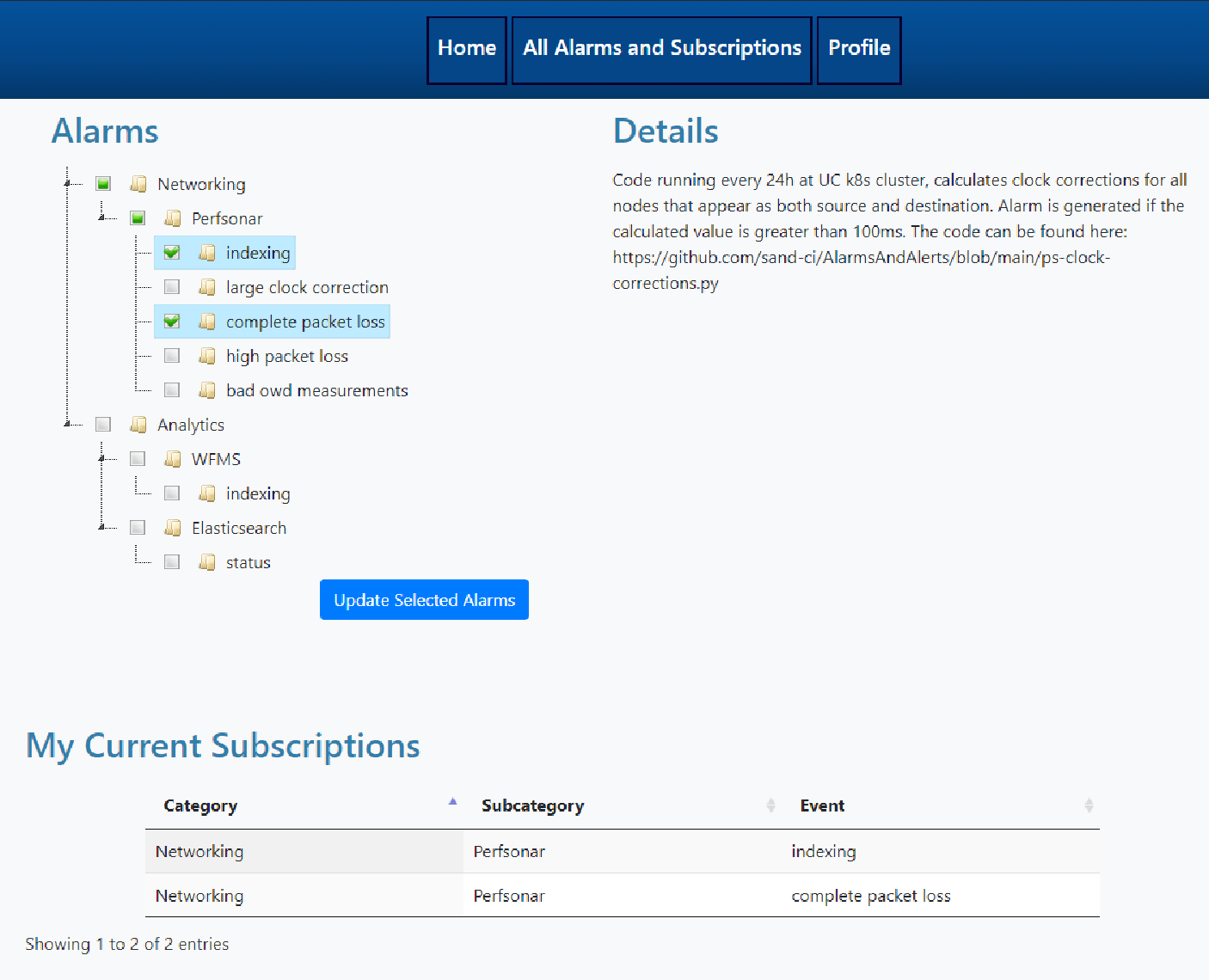}}
\caption{The draft web interface to allow users to self-subscribe to alerts.} 
\label{fig:alert-subscribe}       
\end{figure}

Regarding alerting, we have another student working on a self-subscription interface that exposes the various alerting functions we have enabled.  The web interfaces works directly with the analysis pipeline and analytics platform to dynamically find and allow subscription to new alerts as they are installed.  The current interface is shown in Fig \ref{fig:alert-subscribe}.

\begin{figure}[ht]
\centering
\fbox{\includegraphics[width=0.45\textwidth]{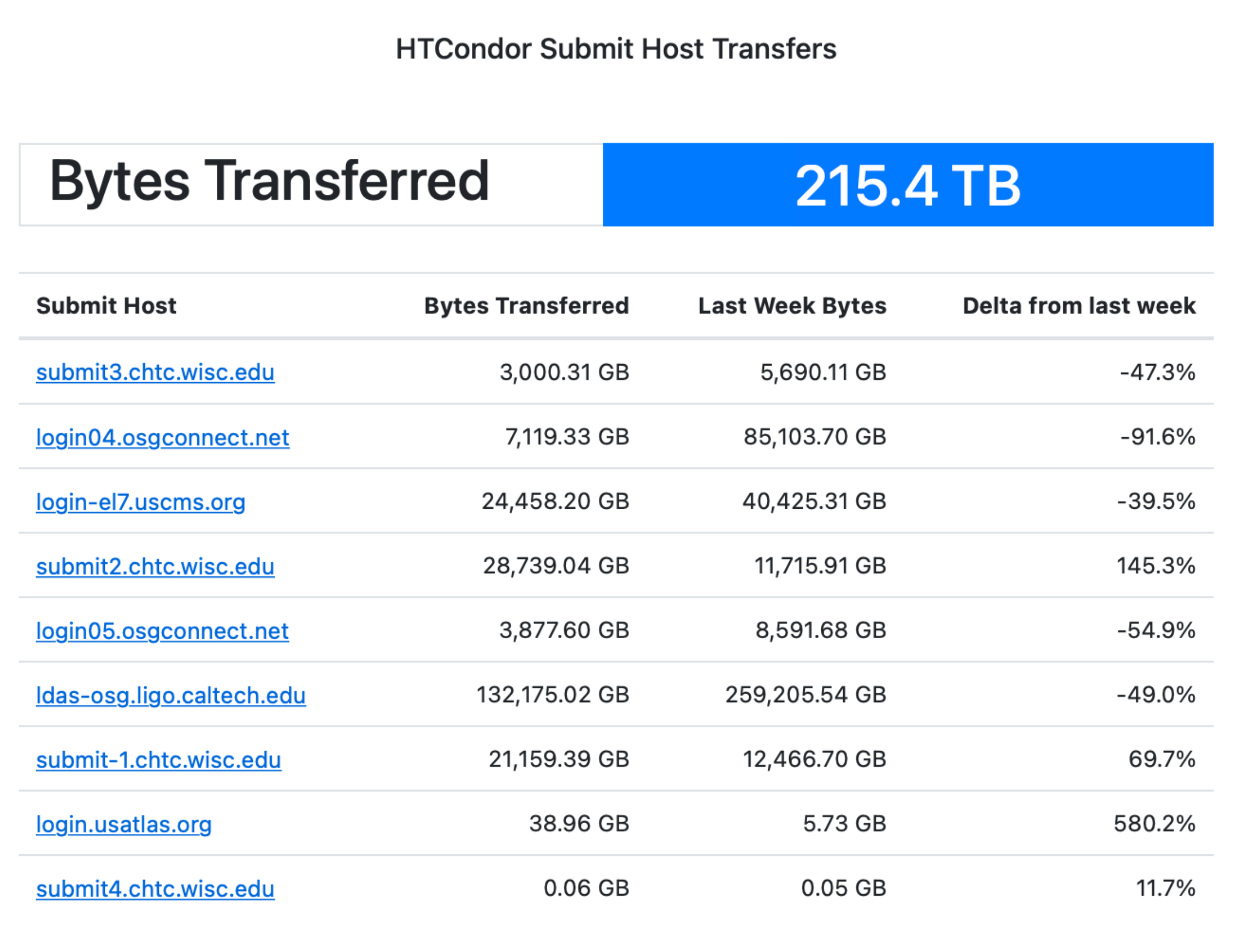}}
\caption{HTCondor TCP metrics email} 
\label{fig:htcondortcpemail}       
\end{figure}

The HTCondor data is visualized through a Grafana dashboard and a weekly email.  A screenshot of the email is shown in Figure \ref{fig:htcondortcpemail}.  This weekly report shows the total amount of data transfers performed by each submit host, the change in volume from the previous week, the average packets lost per transfer, and any hosts that stopped reporting in the week.

\section{OSG and WLCG Network Monitoring Platform}

\begin{figure*}[t]
\centering
\includegraphics[width=\textwidth,height=0.90\textheight,keepaspectratio]{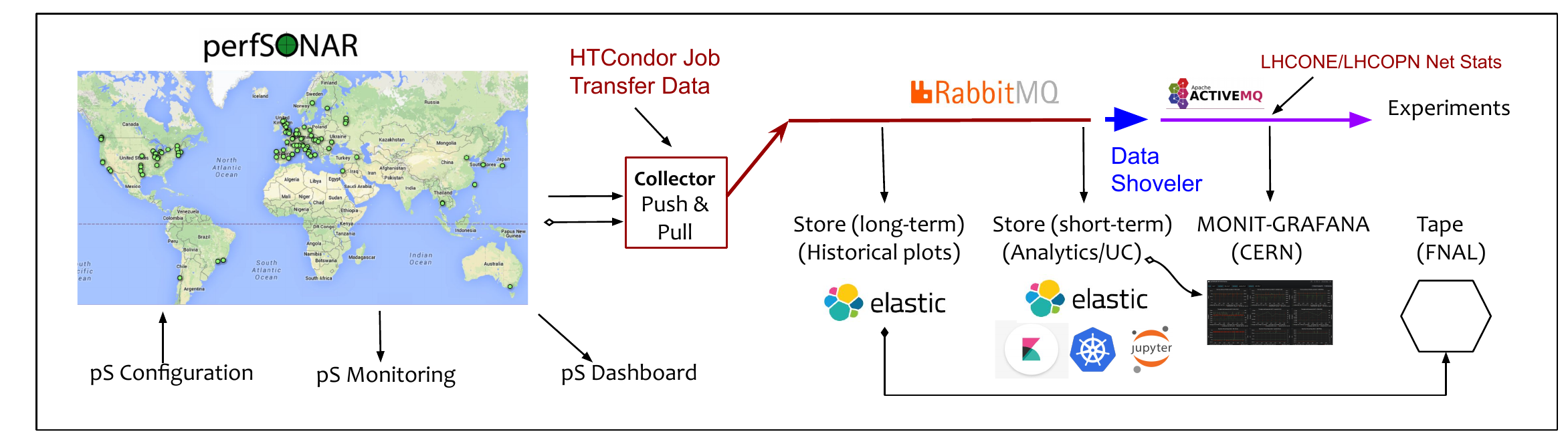}
\caption{OSG Network Monitoring Platform - distributed deployment that collects, stores, visualises and provides APIs for the measurements collected by the WLCG perfSONAR infrastructure} 
\label{fig-4}       
\vspace*{-15pt}
\end{figure*}

The SAND data pipeline is the core of global network monitoring platform \cite{osg-datastore} developed and maintained by OSG and WLCG.  This comprehensive network monitoring platform manages, collects, stores, visualises and further processes all the measurements taken by the by our global perfSONAR deployment.  Figure \ref{fig-4} outlines the entire platform, including the components at CERN developed outside SAND.


The measurement data stream is also available to the LHC experiments at CERN via ActiveMQ bus which is populated by a dedicated bridge (a data ``shoveler") connected directly to RabbitMQ. The platform is also integrated with the ATLAS Analytics and Machine Learning Platform\cite{atlas_analytics} which makes it easy to combine and analyze network measurements with metrics from various different sources (including job management, file transfer, and data management systems).

\label{sec-1}





The global platform also has  infrastructure monitoring\cite{etf,psetf} that oversees the status of the platform and measurement infrastructure and a set of MaDDash \cite{psmad} dashboards that visualize the measurement results.

\section{Platform Use}
\label{sec:platform_use}
Table \ref{tab:sanddbusage} shows a summary of the data volume which flows through the SAND data pipeline. During the three years of operation, the platform has accumulated significant volume of measurements that can provide unique insights into research and education network performance.

\begin{table}[ht]
\vspace{0.2cm}
\centering
\begin{tabular}{c|c|c|c|c}
 &  & \textbf{Total} & \tabularnewline[-2pt]
 & \textbf{Type} & \textbf{Tests} & \textbf{Tests/day} & \textbf{Storage}\tabularnewline
\hline 
 & Latency & 6.91B & 7.95M & 3.1TB\tabularnewline
\cline{2-5}
 & Packet loss & 7.00B & 8.08M & 2.4TB\tabularnewline
\cline{2-5}
\textbf{perfSONAR} & Retransmits & 14.7M & 18.8k & 6.3GB\tabularnewline
\cline{2-5}
 & Throughput & 15.6M & 19.2k & 7.0GB\tabularnewline
\cline{2-5}
 & Network path & 1.28B & 2.14M & 1.5TB\tabularnewline
\hline 
\textbf{ESnet} & Traffic & 1.1B & 44.7M & 1.74TB\tabularnewline
\cline{2-5}
 & Interfaces & 3.2M & 11.8k & 530MB\tabularnewline
\hline 
\textbf{HTCondor} & Job transfers & 734M & 446k & 610GB\tabularnewline
\hline 
\multicolumn{1}{c}{\textbf{Total}} & & 17.1B & 64.4M & 9.4TB\tabularnewline
\end{tabular}
\caption{Storage and incoming rates by data type}
\label{tab:sanddbusage}
\end{table}

The platform and measurement infrastructure have been used in number of activities and collaborations, improving the understanding of the networks and contributing to their technical evolution and design.

Establishing end-site network throughput support has helped the WLCG Network Throughput Working Group to resolve number of challenging cases that would otherwise be difficult to detect and isolate or would take considerable amount of time to resolution\cite{net_cases}. In addition, the working group has helped sites with their data center network design, consulting on the potential bottlenecks caused by the network equipment with insufficient buffers as well as helping to test and benchmark performance. The feedback gathered from the support unit on the different cases has lead to a discussion and a concrete proposal for MTU recommendations for the LHCOPN and LHCONE networks\cite{mtu_doc}, which aims to improve the overall throughput and standardise MTU deployment across R\&Es and sites. 

There were number of significant contributions to the development and design of network performance monitoring over the years; a notable example is the the current configuration system which was initially developed as an internal OSG tool and was later adopted by the perfSONAR consortium.
Another area of close collaboration was deployment and testing of the IPv6 readiness which was led by the HEPiX IPv6 working group\cite{ipv6}. This was a particular example how the platform can be useful in the future to evaluate potential deployment of the new technologies (such as a new TCP congestion control algorithms or software defined networks). Another example is a collaboration with HELIX NEBULA Science Cloud; the project used the platform to assess network performance of the cloud providers. Finally, the SAND team is working to establish close collaborations with other research domains and institutes that have also shown interest in network performance and deployment of a similar platform as the one deployed for OSG and WLCG. 


\section{Network Analytics}
\label{sec:network_analytics}

The \emph{OSG Network Monitoring Platform} is a combination of the SAND data pipeline, perfSONAR toolits, analytics platform, configuration management with PWA, and monitoring. Making the data available for experiments and network researchers has triggered interest from research communities that have started to look at the existing measurements and perform data analysis. The platform has made it possible to diagnose and debug existing network issues, identify the problematic links or equipment and help fix the underlying problems. Analytics projects that have delivered notable results or identified important areas of research include:
\begin{itemize}
\item Real-time detection of “obvious” issues and corresponding alerting and notifications is being developed by students at the University of Michigan and the University of Plovdiv using the ATLAS Analytics and Machine Learning Platform\cite{atlas_analytics}.  A self-subscribe web interface will allow users to select various conditions (e.g., large packet loss or low throughput) that will result in email notification.
\item A study to derive how LHCOPN network paths perform from the existing OWAMP measurements has shown that OWAMP is sufficiently sensitive to pinpoint when network equipment gets stressed and could be used to easily detect peak periods. The main challenge that still remains is how to extend the model to the larger LHCONE network, mainly due to the lack of reliable network traffic data that could be used to train the neural network\cite{babik_borras}. 
\item A visualisation platform\cite{etretyakov_2020_3903716} for network paths was developed in collaboration with MEPhI\footnote{\url{https://eng.mephi.ru/}} which allows selecting and visualising the measured paths between two endpoints. 
\item A network path analysis project is currently ongoing at University of Michigan and aims to calculate simple statistics from the existing path measurements in order to auto-detect potential routing problems and help with the visualisation of the measurements.
\item A study to understand the differences between network utilization as seen by R\&E networks as computed from the experiments data transfers is another area of interest. While there has been significant effort contributed to understand network utilization from the bulk data transfers, there are still major gaps in getting reliable sources of information directly from the R\&E networks.
\end{itemize}

Further analytical studies are planned to better understand our use of networks and how it could be improved. As new versions of perfSONAR are deployed globally and configured to better integrate push-based measurements publishing, we are able to make progress in projects that requiring access to real-time data for automated debugging and optimisations.

\section{Evolution and Future}
The SAND data pipeline provides a mechanism to aggregate and disseminate end-to-end network performance monitoring to communities of interest.  It has been constructed by the NSF-funded SAND project in collaboration with the OSG and WLCG communities and is part of a global network monitoring platform.  The pipeline builds on preexisting foundations, such as the mesh of perfSONAR toolkit endpoints managed by the OSG and WLCG as part of their joint Throughput Monitoring Working Group, and strives to provide a global snapshot of the achievable performance of R\&E networks.

The pipeline architecture is in line with the team's vision; however, we desire to expand the number of application-level network performance data sources.  We are also interested in having the HTCondor data transfer information being posted directly to the RabbitMQ bus instead of to LogStash.  Finally, the push-based perfSONAR pipeline currently distributes JWTs using host-based authentication; this distribution should be transitioned to a stronger authentication method.



Billions of network performance test results have been moved across this pipeline, providing a rich dataset for operations teams.  However, billions of datapoints does not immediately transform into knowledge: as outlined in Sections \ref{sec:platform_use} and \ref{sec:network_analytics}, there are active efforts to build analyses and visualization tools on top of the pipeline.


Over the next year, the Service Analysis and Network Diagnosis (SAND) will be focusing on combining, visualising, and analyzing disparate
network monitoring and service logging data. It will extend and augment the OSG networking efforts with a primary goal of extracting useful insights and metrics from the wealth of network data being gathered from perfSONAR, file transfer services, R\&E network flows and related network information from HTCondor and other applications.


\section*{Acknowledgment}
We gratefully acknowledge the National Science Foundation which supported this work through NSF grants OAC-1836650 and OAC-1827116.   In addition, we acknowledge our collaborations with the CERN IT, WLCG and LHCONE/LHCOPN communities who also participated in this effort.

%
\bibliographystyle{IEEEtran}
\bibliography{IEEEabrv,bibliography}

%
%

\end{document}